\documentclass[]{spie}  

\usepackage{amsmath,amsfonts,amssymb}
\usepackage{graphicx}
\usepackage[colorlinks=true, allcolors=blue]{hyperref}
\usepackage[numbers]{natbib}
\usepackage{booktabs}

\usepackage{paralist}

\title{The HOSTS Survey for Exozodiacal Dust: Preliminary results and future prospects}

\author[a]{Ertel,~S.}
\author[b,c]{Kennedy,~G.~M. }
\author[d]{Defr\`ere,~D.}
\author[a]{Hinz,~P.}
\author[e,f]{Shannon,~A.~B.}
\author[g]{Mennesson,~B.}
\author[h]{Danchi,~W.~C.}
\author[g]{Gelino,~C.}
\author[i]{Hill,~J.~M.}
\author[a]{Hoffmann,~W.~F.}
\author[a]{Rieke,~G.}
\author[a]{Spalding,~E.}
\author[a]{Stone,~J.~M.\footnote{~~Hubble fellow.}~}
\author[a]{Vaz,~A.}
\author[j]{Weinberger,~A.~J.}
\author[g]{Willems,~P.}
\author[d]{Absil,~O.}
\author[a]{Arbo,~P.}
\author[g]{Bailey,~V.~P.}
\author[k]{Beichman,~C.}
\author[g]{Bryden,~G.}
\author[a]{Downey,~E.~C.}
\author[a]{Durney,~O.}
\author[l]{Esposito,~S.}
\author[a]{Gaspar,~A.}
\author[a]{Grenz,~P.}
\author[m]{Haniff,~C.~A.}
\author[a]{Leisenring,~J.~M.}
\author[d]{Marion,~L.}
\author[a]{McMahon,~T.~J.}
\author[k]{Millan-Gabet,~R.}
\author[a]{Montoya,~M.}
\author[a]{Morzinski,~K.~M.}
\author[l]{Pinna,~E.}
\author[i]{Power,~J.}
\author[l]{Puglisi,~A.}
\author[h]{Roberge,~A.}
\author[g]{Serabyn,~E.}
\author[n]{Skemer,~A.~J.}
\author[g]{Stapelfeldt,~K.}
\author[a]{Su,~K.~Y.~L.}
\author[a]{Vaitheeswaran,~V.}
\author[o]{Wyatt,~M.~C.}

\affil[a]{Steward Observatory, Department of Astronomy, University of Arizona, 993 N. Cherry Ave, Tucson, AZ, 85721, USA}
\affil[b]{Department of Physics, University of Warwick, Gibbet Hill Road, Coventry CV4 7AL, UK}
\affil[c]{Centre for Exoplanets and Habitability, University of Warwick, Gibbet Hill Road, Coventry, CV4 7AL, UK}
\affil[d]{Space sciences, Technologies \& Astrophysics Research (STAR) Institute, University of Li\`ege, Li\`ege, Belgium}
\affil[e]{Department of Astronomy and Astrophysics, The Pennsylvania State University, State College, PA 16801, USA}
\affil[f]{Center for Exoplanets and Habitable Worlds, The Pennsylvania State University, State College, PA 16802, USA}
\affil[g]{Jet Propulsion Laboratory, California Institute of Technology, 4800 Oak Grove Dr., Pasadena, CA 91109, USA}
\affil[h]{NASA Goddard Space Flight Center, Exoplanets \& Stellar Astrophysics Laboratory, Code 667, Greenbelt, MD 20771,
  USA}
\affil[i]{Large Binocular Telescope Observatory, University of Arizona, 933 N. Cherry Avenue, Tucson, AZ 85721, USA}
\affil[j]{Department of Terrestrial Magnetism, Carnegie Institution of Washington, 5241 Broad Branch Road NW, Washington, DC,
  20015, USA}
\affil[k]{NASA Exoplanet Science Institute, MS 100-22, California Institute of Technology, Pasadena, CA 91125, USA}
\affil[l]{INAF-Osservatorio Astrofisico di Arcetri, Largo E. Fermi 5, I-50125 Firenze, Italy}
\affil[m]{Cavendish Laboratory, University of Cambridge, JJ Thomson Avenue, Cambridge CB3 0HE, UK}
\affil[n]{Astronomy Department, University of California Santa Cruz, 1156 High Street, Santa Cruz, CA 95064, USA}
\affil[o]{Institute of Astronomy, University of Cambridge, Madingley Road, Cambridge CB3 0HA, UK}

\authorinfo{Send correspondence to Steve Ertel, E-mail: sertel@email.arizona.edu}

\begin{document} 
\maketitle

\pagebreak

\begin{abstract}
The presence of large amounts of dust in the habitable zones of nearby stars is a significant obstacle for future exo-Earth
imaging missions. We executed the HOSTS (Hunt for Observable Signatures of Terrestrial Systems) survey to determine the
typical amount of such exozodiacal dust around a sample of nearby main sequence stars. The majority of the data have been
analyzed and we present here an update of our ongoing work. Nulling interferometry in $N$~band was used to suppress the
bright stellar light and to detect faint, extended circumstellar dust emission. We find seven new $N$~band excesses in
addition to the high confidence confirmation of three that were
previously known. We find the first detections around Sun-like stars and around stars without previously known circumstellar
dust. Our overall detection rate is 23\%. The inferred occurrence rate is comparable for early type and Sun-like stars, but
decreases from 71$^{+11}_{-20}$\% for stars with previously detected mid- to far-infrared excess to 11$^{+9}_{-4}$\% for
stars without such excess, confirming earlier results at high confidence. For completed observations on individual stars, our
sensitivity is five to ten times better than previous results. Assuming a lognormal luminosity function of the dust, we find
upper limits on the median dust level around all stars without previously known mid to far infrared excess of 11.5~zodis at
95\% confidence level. The corresponding upper limit for Sun-like stars is 16~zodis. An LBTI vetted target list of Sun-like
stars for exo-Earth imaging would have a corresponding limit of 7.5~zodis. We provide important new insights into the
occurrence rate and typical levels of habitable zone dust around main sequence stars. Exploiting the full range of
capabilities of the LBTI provides a critical opportunity for the detailed characterization of a sample of exozodiacal dust
disks to understand the origin, distribution, and properties of the dust.  
\end{abstract}


\section{Introduction} \label{sect_intro}
The presence of warm and hot dust in nearby planetary systems both provides important opportunities to study their
architecture and dynamics and impairs our ability to directly image and characterize Earth-like planets. At temperatures
between a few 100\,K to $\sim$2000\,K, such exozodiacal dust clouds (exozodis for short) are located near the habitable zones
(HZs, \cite{mil11, men14, ert18}) of their host stars and closer in \cite{abs06, abs13, ert14b, ert16}. For a recent review
see \cite{kra17}.

The dust may form from comet evaporation \cite{far17}, local asteroid collisions (for young systems and at relatively large
separations from the star, \cite{bac93}), and collisions of large bodies further out in the system from where it migrates
closer to the star due to Poynting-Robertson and stellar wind drag \cite{ken15b, rei11}. It gravitationally interacts with
any planets in the system in addition to the star and is ground to (sub-)micron sizes through mutual collisions of dust
grains, after which the stellar radiation and wind pressure cause the grains to be expelled from the system. The dust 
distribution may trace the location of the dust producing planetesimals or in case of cometary origin the magnitude of the
cometary activity. Structures in the dust distribution such as gaps and resonant clumps can reveal the presence and
parameters of planets \cite{ert12b, sha15}.

At the same time, bright dust clouds constitute the main source of confusion and noise for future exo-Earth imaging
missions. The mere presence of dust emission (scattered light or thermal emission may dominate depending on the wavelength of
observation) would add photon noise to the observations, even if the emission itself could be perfectly removed. Moreover,
dust clumps due to resonant trapping by a planet can be misinterpreted as actual planets due to the limited angular
resolution and signal-to-noise ratio of the observations. Thus, the expected amount of dust present around the target stars
determines the design of any mission to detect exo-Earths, such as the required angular resolution and size of the primary mirror (e.g., \cite{sta15, sta16, def10}).

We have conducted the NASA funded HOSTS survey (Hunt for Observable Signatures of Terrestrial Systems) to search for
habitable zone dust around a sample of nearby stars and to measure their dust levels. We observed in $N$~band, where the flux
ratio between warm ($\sim$300\,K) dust and the star is most favorable while the atmosphere is still sufficiently transparent.
The goal was to detect HZ dust levels only a few times more massive than the Solar system level, corresponding to an $N$~band
excess emission of the order of $10^{-5}$. We thus used nulling interferometry at the Large Binocular Telescope
Interferometer (LBTI, \cite{hin16}) in order to spatially disentangle the dust emission from that of the star and to suppress
the bright star light.

The survey observations have now been completed, but a fraction of the newest observations are still being analyzed. We
discuss the latest results of our ongoing work and future prospects for the characterization of exozodiacal dust with the
LBTI.

\section{Observations and Data Analysis}

A detailed description of our observations and data reduction can be found in \cite{ert18}. We present here only a brief
summary.

Target stars have been selected for observations in real time from the full HOSTS target list presented by \cite{wei15}. Our
target stars are bright, nearby main sequence stars ($F_{\nu}>1$\,Jy in $N$~band) without known close ($<1.5''$) binary
companions. Spectral types range from A0 to K8. A list of observed stars with basic, relevant properties is given in
Table~\ref{tab_targets}. Calibrators were selected following \cite{men14} using the catalogs of \cite{bor02} and
\cite{mer05}, supplemented by stars from the JSDC catalog and the SearchCal tool (both \cite{che16}) where necessary.

\begin{table}[]
\caption{Observed stars included in this work.} 
\label{tab_targets}
{\footnotesize
\begin{tabular}{ccccccccccc}
\hline\hline
 HD & Name & \# SCI$^{~a}$ & Sp.\ Type & $V$   & $K$   & $N'$      & $d$  & EEID        & fIR/nIR & Excess     \\
 number &  &               &           & (mag) & (mag) & (Jy)      & (pc) & (mas)       & excess  & references \\
\hline
 \multicolumn{11}{l}{\textbf{Sensitivity driven sample}\textbf{:}}\\
\hline
  33111   & $\beta$\,Eri   & 2  & A3\,IV      & 2.782  & 2.38  & 3.7  & 27.4  & 248  & N/N      &
\cite{gas13, thu14, ert14b} \\
  *38678  & $\zeta$\,Lep   & 1  & A2\,IV--V   & 3.536  & 3.31  & 2.1  & 21.6  & 176  & Y/N      &
\cite{gas13, ert14b}        \\
  *81937  & 23\,UMa        & 5  & F0\,IV      & 3.644  & 2.73  & 2.6  & 23.8  & 168  & N/--     &
\cite{bei06}                \\
  95418   & $\beta$\,UMa   & 4  & A1\,IV      & 2.341  & 2.38  & 4.2  & 24.5  & 316  & Y/N      &
\cite{su06, abs13}          \\
  97603   & $\delta$\,Leo  & 4  & A5\,IV      & 2.549  & 2.26  & 3.9  & 17.9  & 278  & N/N      &
\cite{gas13, thu14, abs13}  \\
  103287  & $\gamma$\,UMa  & 4  & A0\,V       & 2.418  & 2.43  & 3.7  & 25.5  & 308  & N/--     &
\cite{gas13, thu14, su06}   \\
  106591  & $\delta$\,UMa  & 4  & A2\,V       & 3.295  & 3.10  & 2.0  & 24.7  & 199  & N/N      &
\cite{gas13, thu14, abs13}  \\
  108767  & $\delta$\,Crv  & 2  & A0\,IV      & 2.953  & 3.05  & 2.3  & 26.6  & 251  & N/Y      &
\cite{gas13, thu14, ert14b} \\
  128167  & $\sigma$\,Boo  & 3  & F4\,V       & 4.467  & 3.47  & 1.4  & 15.8  & 117  & Y/N      &
\cite{gas13, abs13}         \\
  129502  & $\mu$\,Vir     & 3  & F2\,V       & 3.865  & 2.89  & 2.6  & 18.3  & 151  & N/N      &
\cite{gas13, thu14}         \\
  *172167 & $\alpha$\,Lyr  & 4  & A0V         & 0.074  & 0.01  & 38.6 & 7.68  & 916  & Y/Y      &
\cite{su06, abs06}          \\
  187642  & $\alpha$\,Aql  & 2  & A7\,V       & 0.866  & 0.22  & 21.6 & 5.13  & 570  & N/Y      &
\cite{gas13, thu14, abs13, rie05} \\
  203280  & $\alpha$\,Cep  & 1  & A8\,V       & 2.456  & 1.85  & 7.0  & 15.0  & 294  & N/Y      &
\cite{gas13, thu14, abs13, rie05} \\
\hline
 \multicolumn{11}{l}{\textbf{Sun like stars sample}\textbf{:}}\\
\hline
  *9826   & $\mu$\,And     & 2  & F9\,V       & 4.093  & 2.85  & 2.4  & 13.5  & 136  & N/N      &
\cite{bei06, abs13, eir13}  \\
  10476   & 107\,Psc       & 3  & K1\,V       & 5.235  & 3.29  & 2.0  & 7.53  & 90   & N/N      &
\cite{gas13, abs13, mon16, tri08} \\
  *10700  & $\tau$\,Cet    & 2  & G8\,V       & 3.489  & 1.68  & 5.4  & 3.65  & 182  & Y/Y      &
\cite{abs13, gre04}         \\
  16160   & GJ\,105\,A     & 1  & K3\,V       & 5.815  & 3.45  & 1.5  & 7.18  & 73   & N/--     &
\cite{gas13, mon16, tri08}  \\
  *22049  & $\epsilon$\,Eri & 4 & K2\,V       & 3.721  & 1.66  & 7.4  & 3.22  & 172  & Y/N      &
\cite{abs13, aum85}         \\
  *30652  & 1\,Ori         & 4  & F6\,V       & 3.183  & 2.08  & 4.8  & 8.07  & 205  & N/N      &
\cite{gas13, abs13, mon16, tri08} \\
  34411   & $\lambda$\,Aur & 2  & G1\,V       & 4.684  & 3.27  & 1.8  & 12.6  & 105  & N/--     &
\cite{tri08, law09}         \\
  48737   & $\xi$\,Gem     & 3  & F5\,IV-V    & 3.336  & 2.13  & 4.3  & 18.0  & 196  & --/N     &
\cite{abs13}                \\
  *78154  & 13\,UMa        & 3  & F7\,V       & 4.809  & 3.53  & 1.2  & 20.4  & 99   & N/--     &
\cite{gas13}                \\
  88230   & GJ\,380        & 2  & K8\,V       & 6.598  & 3.21  & 1.9  & 4.87  & 65   & N/--     &
\cite{eir13}                \\
  89449   & 40\,Leo        & 2  & F6\,IV-V    & 4.777  & 3.65  & 1.1  & 21.4  & 98   & N/--     &
\cite{gas13, bei06}         \\
  *102870 & $\beta$\,Vir   & 4  & F9\,V       & 3.589  & 2.32  & 4.3  & 10.9  & 173  & N/N      &
\cite{abs13, tri08}         \\
  *120136  & $\tau$\,Boo   & 4  & F6\,IV      & 4.480  & 3.36  & 1.7  & 15.6  & 114  & N/N      &
\cite{ert14b, tri08, law09} \\
  126660  & $\theta$\,Boo  & 3  & F7\,V       & 4.040  & 2.81  & 3.1  & 14.5  & 147  & N/--     &
\cite{gas13, mon16, tri08}  \\
  141004  & $\lambda$\,Ser & 2  & G0\,IV-V    & 4.413  & 2.98  & 2.4  & 12.1  & 121  & N/N      &
\cite{gas13, abs13, mon16, koe10} \\
  142373  & $\chi$\,Her    & 3  & G0\,V       & 4.605  & 3.12  & 2.0  & 15.9  & 111  & N/N      &
\cite{gas13, bei06, abs13, mon16} \\
  142860  & $\gamma$\,Ser  & 4  & F6\,IV      & 3.828  & 2.63  & 2.9  & 11.3  & 151  & N/N      &
\cite{gas13, abs13, mon16, law09} \\
  *173667 & 110\,Her       & 5  & F6\,V       & 4.202  & 3.03  & 2.2  & 19.2  & 131  & Y/Y      &
\cite{abs13, eir13}         \\
  185144  & $\sigma$\,Dra  & 2  & G9\,V       & 4.664  & 2.83  & 2.7  & 5.76  & 113  & N/N      &
\cite{abs13, tri08, law09}  \\
  215648  & $\xi$\,Peg\,A  & 3  & F6\,V       & 4.203  & 2.90  & 2.2  & 16.3  & 132  & N/N      &
\cite{gas13, bei06, mon16}  \\
  *222368 & $\iota$\,Psc   & 2  & F7\,V       & 4.126  & 2.83  & 2.4  & 13.7  & 137  & N/--     &
\cite{gas13, bei06, mon16}  \\
\hline
\multicolumn{11}{l}{\textbf{Commissioning targets:}}\\
\hline
  102647  & $\beta$\,Leo   & 2  & A3\,V       & 2.121  & 1.92  & 6.9  & 11.0  & 336  & Y/Y      &
\cite{su06, abs13}          \\
  109085  & $\eta$\,Crv    & 3  & F2\,V       & 4.302  & 3.54  & 1.8  & 18.3  & 125  & Y/N      &
\cite{abs13, aum88}         \\
\hline
\end{tabular}}\smallskip\\
The full list of stars observed by the HOSTS survey and all null measurements and derived zodi levels will be published in a
dedicated paper (Ertel et al., in prep.). Stars with new observations since \cite{ert18} are marked with an asterisk. EEID is
the Earth Equivalent Insolation Distance, the distance at which a body receives the same flux density from its host star as
Earth from the Sun. Magnitudes are given in the Vega system.\\
$^a$~Number of calibrated science pointings obtained.
References are:
Spectral type: SIMBAD;
$V$~magnitude: \cite{kha07};
$K$~magnitude: \cite{gez93} and the Lausanne photometric data base (http://obswww.unige.ch/gcpd/gcpd.html);
$N$~band flux and EEID: \cite{wei15};
Distance: \cite{vanle07}\\
\end{table}

\begin{figure}[t]
 \centering
 \includegraphics[width=0.8\linewidth]{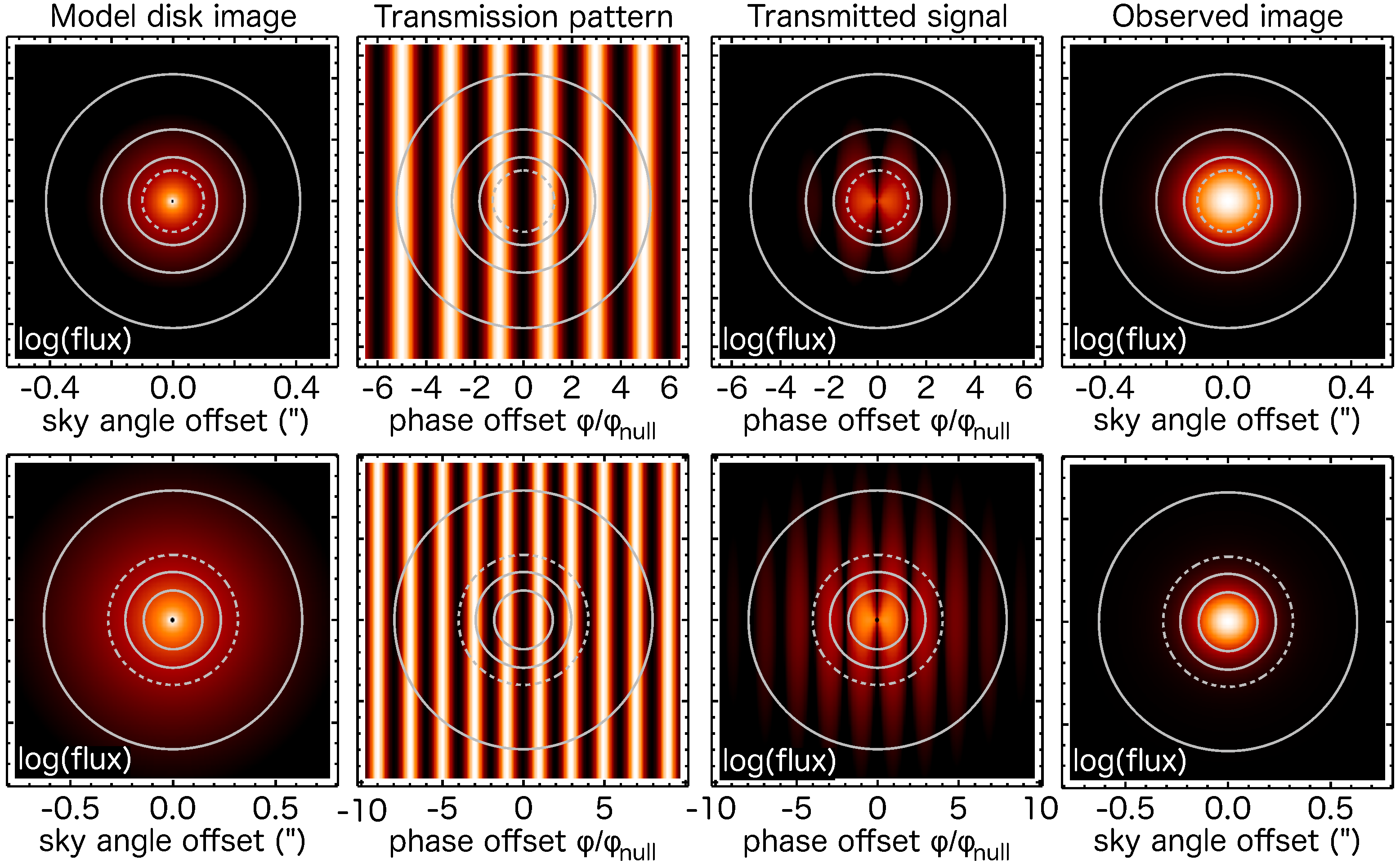}
 \caption{Illustration of the data obtained by nulling interferometry at the LBTI. Model observations are shown for two
  example stars (\emph{top row}: 1\,L$_\odot$, \emph{bottom row}: 10\,L$_\odot$) at a distance of 10\,pc. The panels show
  \emph{from left to right}: A model image of a face-on disk, the LBTI transmission pattern, the transmission pattern applied
  to the model image, and the final simulated observation after convolving with the single aperture PSF. The disk models
  with and without the transmission pattern applied are shown in logarithmic scale, the other images are shown in linear
  scale. The dashed circle marks the size of the HZ (distance from the star receiving the same flux density as Earth from the
  Sun). The three solid circles mark three different photometric apertures used for the HOSTS survey. Simulations are shown,
  because this is typically not visible in real data due to various noise terms that can only be removed by a statistical
  analysis.}
 \label{fig_transmission}
\end{figure}

The HOSTS survey was carried out with the LBTI on the Large Binocular Telescope (LBT) on Mt. Graham, Arizona. The wavefronts
from the two 8.4\,m apertures -- separated by 14.4\,m center to center -- are stabilized using adaptive secondary mirrors.
They are then brought to destructive interference in the LBTI and imaged on the Nulling Optimized Mid-Infrared Camera
(NOMIC, \cite{hof14}). Tip-tilt and optical path delay (OPD) tracking are performed using PhaseCam, a closed loop fringe
tracker \cite{def15b}. The adaptive optics system operates in the $R$~to $I$~band range, PhaseCam in the $K$~band, and
nulling data are recorded in the $N$~band. The interferometric inner working angle is 70\,mas. This is illustrated in
Fig.~\ref{fig_transmission}. A nodding sequence is performed for sky subtraction and photometric observations of each star
are obtained for relative flux calibration. Science observations (SCI) are paired with identical calibrator observations
(CAL) to determine the nulling interferometric transfer function (TF) of the instrument.

After a basic reduction of each frame (nod subtraction, bad pixel correction), photometry is performed on each single frame.
The raw null depth and its uncertainty are determined using the null self calibration (NSC, \cite{men11, han11, def16,
men16a}). Null measurements are converted to dust levels using a power-law radial dust distribution model analogous to the
Solar system's zodiacal dust \cite{kel98, ken15}. The vertical optical depth of the dust at a separation of 1\,AU from the
Sun of $7.12 \times 10^{-8}$ defines our unit of 1\,zodi.

The beam combination in the $N$ band is achieved in the pupil plane before the light is re-imaged in the focal plane. A
consequence is that the light from a point source has the same phase over the entire detector, but the phase is different at
different sky angles. This results in a spatial filtering in form of a transmission pattern (stripes of high and low
transmission vertical to the baseline orientation) being applied to the flux distribution on sky
(Fig.~\ref{fig_transmission}).

\section{Results}
\label{sect_results}

\subsection{Excess significance and detection threshold}
\label{sect_significance}
\begin{figure}[t]
 \centering
 \includegraphics[width=\textwidth]{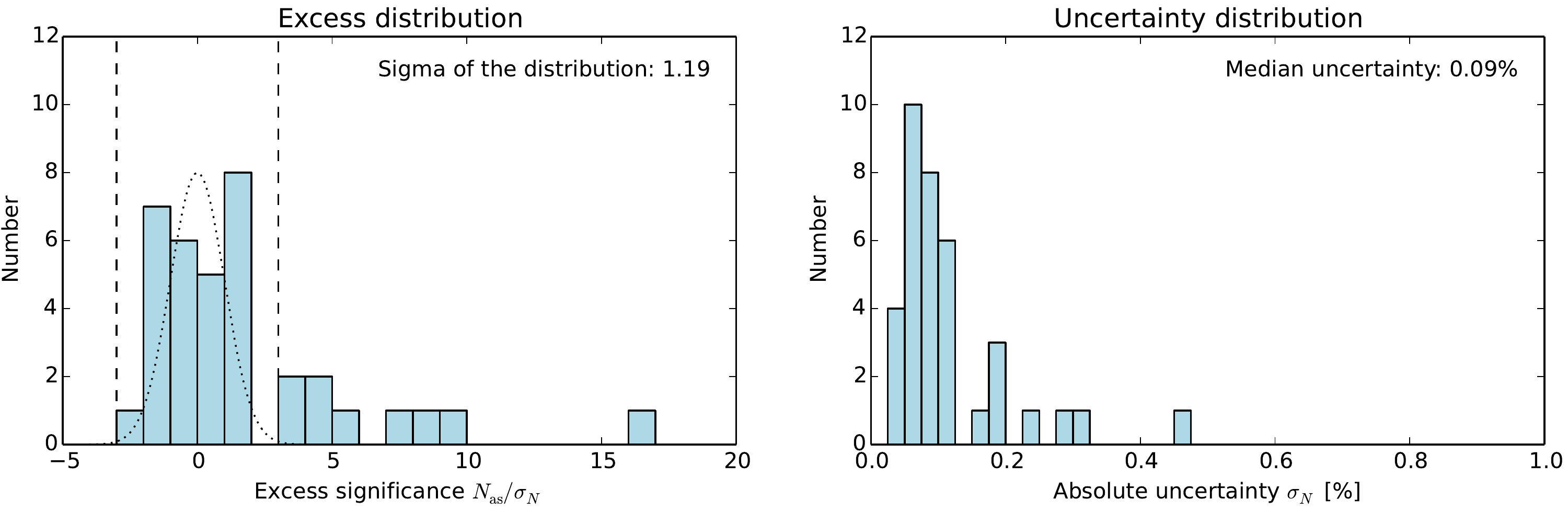}
 \caption{Distribution of excess significance $N_{\text{as}} / \sigma_N$ (\emph{left}) and uncertainties $\sigma_N$
  (\emph{right}). The two vertical, dashed lines in the excess significance distribution plots mark the $\pm 3\,\sigma$
  boundaries based on our uncertainty estimates. The standard deviation of the distribution is computed from non-detections
  only ($-3 < N_{\text{as}} / \sigma_N < 3$). The dotted line represents a Gaussian with a standard deviation of one (Normal
  distribution) scaled to the peak of the histogram and is used only to guide the eye.}
 \label{fig_nulls_errors}
\end{figure}

Fig.~\ref{fig_nulls_errors} shows the distributions of the excess significance $N_{\text{as}} / \sigma_N$ and of the
uncertainties. The distributions are in general well behaved with $N_{\text{as}} / \sigma_N$ following a Normal distribution
for $-3 < N_{\text{as}} / \sigma_N < 3$ with the addition of several detections at $N_{\text{as}} / \sigma_N > 3$. We apply a
3$\,\sigma$ threshold to identify significant excesses. We find ten significant excess detections.

\subsection{Updates on specific targets}
\label{sect_res_spec}

Seven of our ten detections have already been discussed in detail by \cite{ert18}. Among those, no new data are included in
this work for $\beta$\,Leo, $\eta$\,Crv, $\beta$\,UMa, $\delta$\,UMa, and $\theta$\,Boo. New data have been taken and the
results are included here for $\epsilon$\,Eri and 110\,Her. Both detections were originally only achieved with a very large
aperture size and at relatively low significance. They thus required follow-up observations. For $\epsilon$\,Eri we confirm
our detection with new, high accuracy data and measure a zodi level of 368$\pm$68~zodis. A detection is also achieved in a
smaller aperture (reduced background and read noise). The detection around 110\,Her is also confirmed with a new measurement
of 343$\pm$57~zodis. This detection is, however, still only achieved in a large aperture, with strong limits for smaller
apertures, suggesting that the dust may be concentrated at a relatively large separation from the star. As discussed already
by \cite{ert18} the low significance detection of far-infrared excess around this star is controversial, so that it is not
clear whether to include it in the group of stars with or without previously detected cold dust.

\begin{figure}[t]
 \centering
 \includegraphics[width=0.4\textwidth]{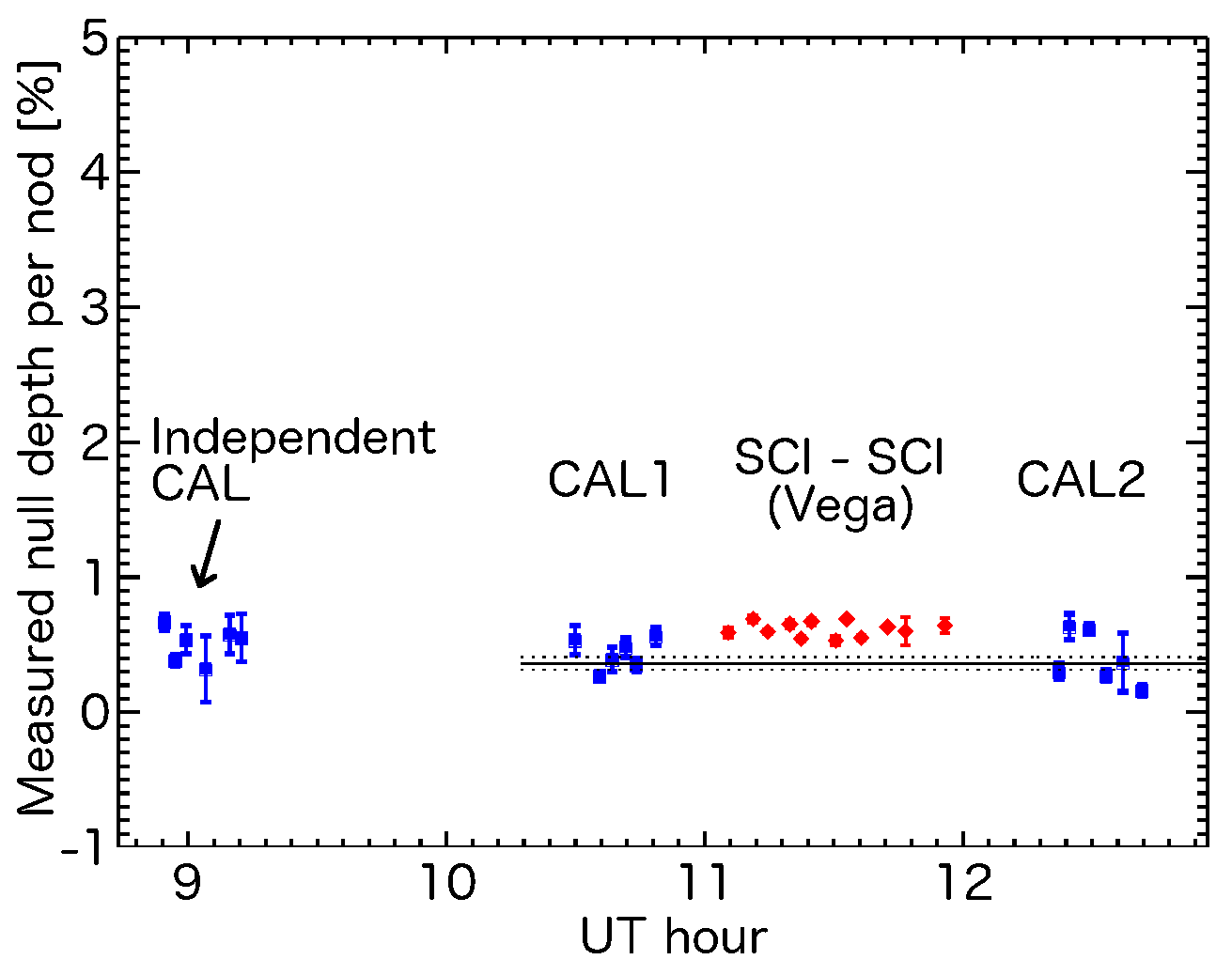}
 \caption{A HOSTS observing sequence (CAL--SCI--SCI--CAL) on Vega showing the faint (4.5$\sigma$ excess). The blue points
  are calibrator observations, the red points are science observations. Measurements from each single nod are displayed. The
  small vertical offset between science and calibrator measurements is the null excess. The first calibrator shown is an
  unrelated observation of a star at a different sky location. The data illustrate the high instrument stability both over
  time and over pointing direction.}
 \label{fig_vega}
\end{figure}

In the new data obtained since summer 2017 (not included in \cite{ert18}), we find three new detections: The Sun-like star
13\,UMa shows a very strong excess of 550$\pm$66~zodis. After $\delta$\,UMa and $\theta$\,Boo, this is the third star in our
survey without previously known cold excess, but with an LBTI detection (potentially the fourth one when including 110\,Her,
see above). This further supports our conclusion in \cite{ert18} that focussing on stars without known cold dust for
exo-Earth imaging does not guarantee a low HZ dust level. We further confirm the detection of HZ dust around $\zeta$\,Lep by
\cite{men14} and measure a zodi level of 739$\pm$46~zodis. 

Most interesting is the fact that new data obtained show a detection of 38$\pm$8\,zodis around Vega. Our new observations
achieved a better sensitivity and the detection is consistent with the previous upper limit presented in \cite{ert18}. The
excess is the strongest in a large aperture and not detected in a very small one, which means we are not seeing the hot dust
in this system previously reported by \cite{abs06}, but rather dust further out, near the HZ. An observing sequence on Vega
is shown in Fig.~\ref{fig_vega}.

Finally, our upper limit on $\tau$\,Cet of 74\,zodis (3$\sigma$) is noteworthy. This star hosts a known debris disk (e.g.,
\cite{law14}), hot dust \cite{abs13}, and four claimed super-Earth mass planets ($\tau$\,Cet~e, f, g, h, \cite{fen17}) with
$\tau$\,Cet~f being in the HZ, but also at sufficient angular separation that it might be detectable with a coronagraph on
WFIRST. Our upper limit shows that despite the high dust content in the system at large and very small separations, the HZ is
relatively dust poor.

\subsection{Statistical results}

In this section, we provide an update of the statistical analysis from \cite{ert18} considering our additional data.

\begin{figure}[t]
 \centering
 \includegraphics[width=0.5\textwidth]{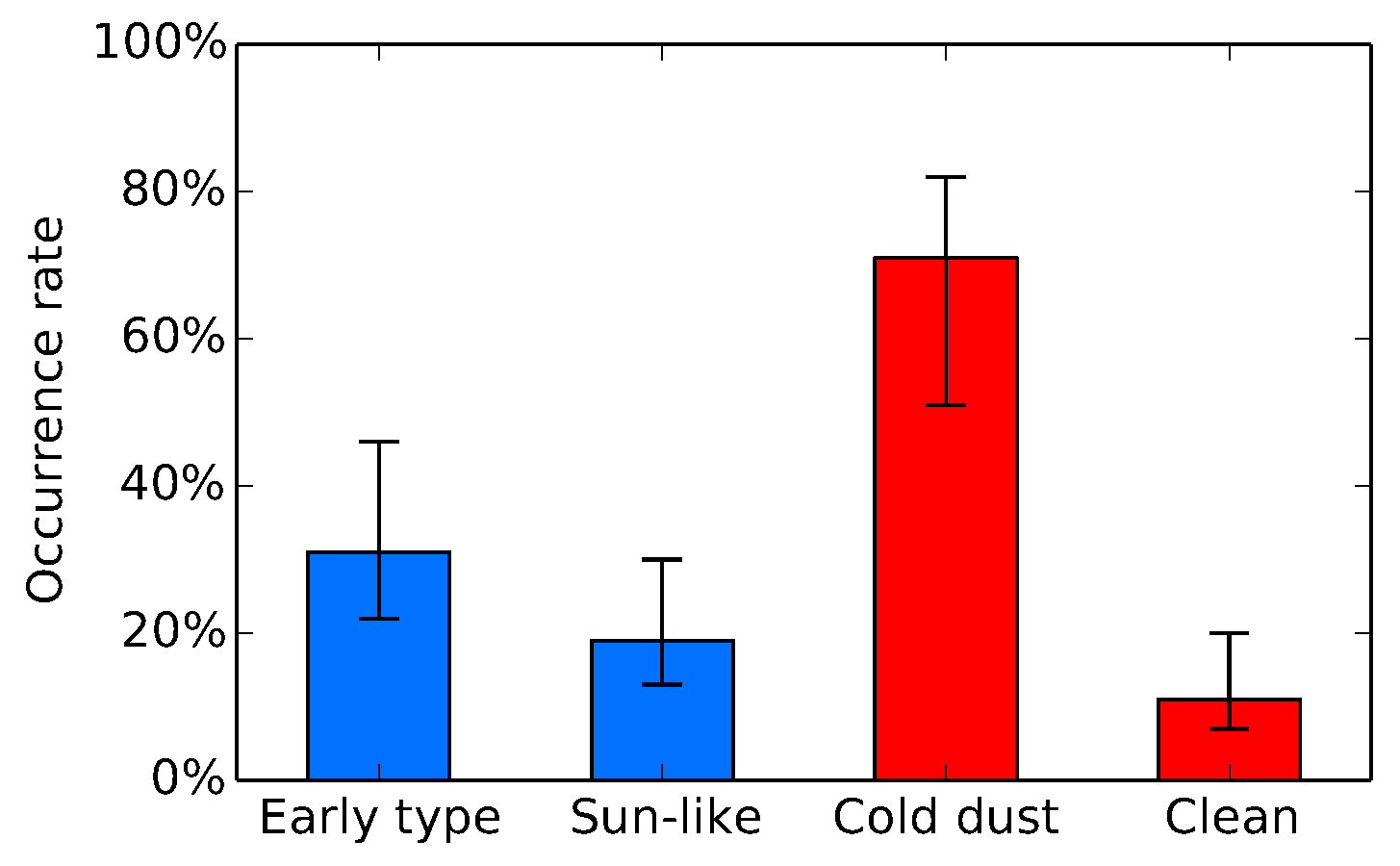}
 \caption{Occurrence rates of exozodiacal dust inferred from the observations considered in this work for early-type stars,
 Sun-like stars, stars with previously known cold dust, and stars without detected cold dust.}
 \label{fig_det_rates}
\end{figure}

We divide our sample into early type stars and Sun-like stars, and into stars with previously known cold dust (`dusty stars')
and without (`clean stars'). As in \cite{ert18} we exclude $\eta$\,Crv and $\beta$\,Leo from the statistical analysis as
biased targets. Detection rates for these subsamples are listed in Table~\ref{tab_det_rates} and shown in
Fig.~\ref{fig_det_rates}. The results of Fisher's exact test to check if differences in the detection rates are significant
are shown in Table~\ref{tab_fisher}. We can rule out with high confidence (probability 0.3\%) that the occurrence rate of
HZ dust is the same among clean and dusty stars. Note that this result is affected by small number statistics of detections
often at the boarder of significance, both in our mid-infrared nulling data and in the far-infrared. For example, moving
110\,Her into the category of stars without a significant cold dust detection, this result moves to a considerably lower
significance probability changes to 0.017. The probability for having the same occurrence rate for stars with and without
known cold dust among Sun-like stars only is nominally 0.08, and becomes 0.32 when considering 110\,Her as a star without
cold dust. We can thus not claim a significant correlation between cold and warm dust for Sun-like stars and the correlation
for all stars needs to be treated with care. The occurrence rates for Sun-like and early-type stars are the same within the
binomial uncertainties. This is surprising given the typically $\sim$4 times lower sensitivity to HZ dust surface density
around Sun-like compared to early-type stars \cite{ken15, wei15}).

Next, we repeat the statistical analysis presented by \cite{ert18} to determine the typical HZ dust levels around our target
stars, including the newest data available. We fit a lognormal exozodi luminosity function with parameters $\mu$ and
$\varsigma$ to our measurements using a maximum likelihood estimate and performing a Bayesian analysis. We apply a $1/m$
prior, equivalent to assuming a flat prior in $\mu$, marginalize the likelihood distribution over $\varsigma$, and compute
the posterior cumulative probability distribution function (CPDF) of $m = \exp\left(\mu\right)$. For Sun-like stars without
previously detected cold dust, we currently find a 95\% confidence upper limit on the median zodi level of 16~zodis. These
stars are the highest priority targets for future exo-Earth imaging missions.

\begin{table}[!t]
~\smallskip\\
\caption{Subsamples, excess detections, and occurrence rates\label{tab_det_rates}}
\centering
\begin{tabular}{cccc}
\hline
\hline
 ~~~~~~~~~~~~ & ~~~Cold dust~~~ & ~~~~~Clean~~~~~ & ~~~~~~All~~~~~~ \\
\hline
 Early                & 3 of 4              & 1 of 9            & 4 of 13            \\[-2pt]
 ~~type               & $75^{+10}_{-27}\%$  & $11^{+18}_{-4}\%$ & $31^{+15}_{-9}\%$  \\[4pt]
 Sun-                 & 2 of 3              & 2 of 18           & 4 of 21            \\[-2pt]
 ~~like               & $67^{+15}_{-28}\%$  & $11^{+12}_{-4}\%$ & $19^{+11}_{-6}\%$  \\[4pt]
 All                  & 5 of 7              & 3 of 27           & 8 of 34            \\[-2pt]
                      & $71^{+11}_{-20}\%$  & $11^{+9}_{-4}\%$  & $23^{+9}_{-6}\%$   \\
\hline
\end{tabular}\bigskip\\
\end{table}

\begin{table}[!t]
\caption{Probability that two samples are drawn from the same distribution\label{tab_fisher}}
\centering
\begin{tabular}{ccc}
\hline
\hline
 ~~~~~~Samples 1~~~~~~ & ~~~~~~Sample 2~~~~~~ & ~~Probability~~ \\
\hline
 All early type   & All Sun-like     & 0.24 \\
 All dusty        & All clean        & 0.003 \\
 Clean early type & Dusty early type & 0.05 \\
 Clean Sun-like   & Dusty Sun-like   & 0.08 \\
 Clean early type & Clean Sun-like   & 0.47 \\
 Dusty early type & Dusty Sun-like   & 0.57 \\
\hline
\end{tabular}
\end{table}

\section{Future prospects}

\subsection{Extending HOSTS and target vetting for future exo-Earth imaging}
This summer 2018, the LBTI will undergo an adaptive optics upgrade designed to improve its wavefront stability and
performance at low target elevation. In addition, we are improving our application of the NSC to improve background
subtraction (currently a dominant source of uncertainty) with minimal change to the data acquisition strategy. With the
improved instrument performance and sensitivity, new survey observations will provide even stronger statistical results. 

Furthermore, our detections around $\theta$\,Boo and 13\,UMa (both Sun-like stars without known cold dust) show that
selecting such stars for exo-Earth imaging does not guarantee low HZ dust levels. One option to improve over this selection
strategy is to perform target vetting with the LBTI and a suitable future instrument in the South \cite{def18, def18b}. To
illustrate the improvement possible with such a strategy, we produce from our sample of Sun-like stars without cold dust an
LBTI vetted sample by excluding the HOSTS detections. For such a sample we find an upper limit on the median zodi level of
7.5~zodis, an improvement by a factor of two. This translates in a 20\% increase in yield for an exo-Earth imaging mission
of a given size \cite{sta15}.

\begin{figure}[t]
 \centering
 \includegraphics[width=\textwidth]{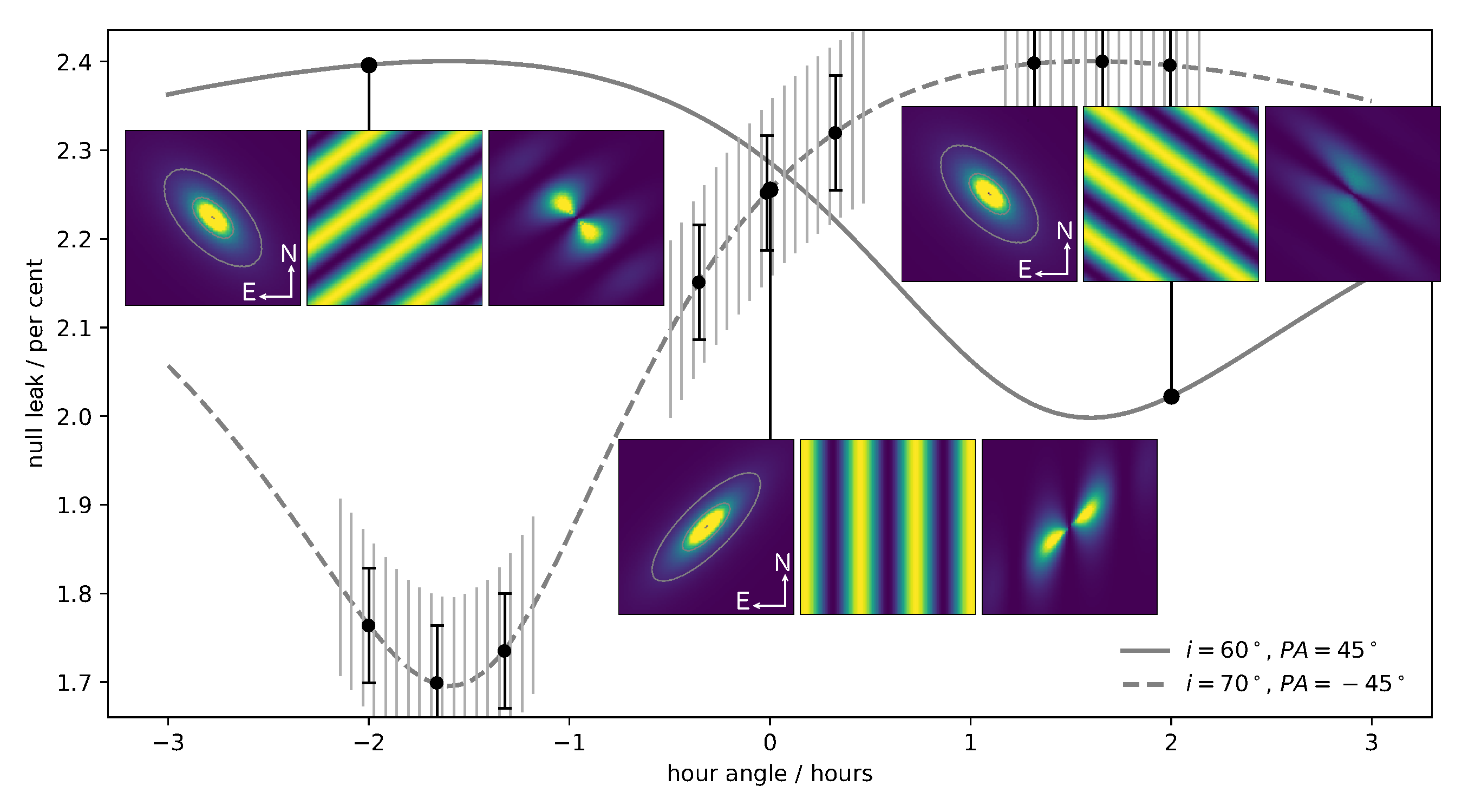}
 \caption{Simulation of observed null leak vs.\ hour angle of observation for two disks with different position angles and
   inclinations. The disk model, the transmission pattern, and the transmitted signal are shown for three representative
   cases. \textit{Over plotted} on the $i=70$\,deg curve is a planned observing sequence and uncertainties for the proposed
   observations. The high cadence, gray bars show the single nods and relative errors (relevant to detect variations over
   HA), while the black points and uncertainties show the measurement when combining 6 nods including absolute calibrations
   (relevant when combining observations from different nights). The large gaps are due to calibrator observations. Our
   detection around $\beta$\,UMa is used as a template.}
 \label{fig_inclinations}
\end{figure}

\subsection{Characterization of detected systems using nulling interferometry}
The performance of LBTI for the HOSTS survey has reached a level that makes nulling infrared interferometry a powerful tool
to study warm dust systems around nearby main sequence stars. With a detection rate of the HOSTS survey of $\sim$25\% and a
sensitivity of only a few zodis for our most favorable stars, we can now study systems with typical HZ dust levels for the
first time. Our high signal-to-noise detections allow for detailed follow-up observations.

The alt-azimuth mount of the LBTI allows us to measure the excess at different position angles of the single interferometric
baseline projected on sky when observing at different hour angles. A different disk size along different position angles (an
inclined disk) or disk structure cause variation of the null excess with baseline orientation. This allows us to measure the
inclination and position angle of the disk, and to search for structures due to planet-disk interaction. The latter will
rotate with the orbital period of the responsible planet, which is of the order of a few years for planets close to the HZ.
Thus, monitoring over a few years can be used to to better distinguish between different scenarios for the hour angle
variation and to constrain the orbit of the responsible planet. In addition, observations at different wavelengths across the
$N$~band provide different spatial filtering by the transmission pattern, thus further constraining the geometry of the
emission. Multi-wavelength data also allow us to constrain the spectral shape of the excess emission.

These information allow us to address the most pressing questions on exozodiacal dust after the HOSTS survey. Determining the
geometry and radial dust distribution is critical for an accurate determination of the zodi level from our nulling data.
Information on the radial dust distribution for a range of systems with different zodi levels and stellar spectral types will
also allow us to make better estimates of the probable levels of exozodi in systems where it has not been detected. This will
allow us to derive more reliable upper limits and thus more reliable statistics from the HOSTS survey. The radial slope of
the dust distribution can inform us about the dust origin: Poynting-Robertson drag from an outer disk would result in a flat
or inwardly decreasing surface density distribution \cite{ken15b}, while a cometary origin would produce an inwardly
increasing one \cite{nes10, mar16}. The detection of potential disk structures and radial discontinuities of the surface
density distribution allows us to put constraints on the responsible planets (orbit, mass, \cite{ert12b, sha15}, Bonsor et
al., in prep.). This provides present-day insight into the architecture and dynamics of nearby planetary systems in and near
their HZs. Information on the spectral slope of the dust emission allows us to constrain the dust grain size \cite{ert12a}
and thus its origin (Poynting-Robertson drag, local collisional production, or cometary origin). Moreover, a better
understanding of the dust grain size allows us to better predict its scattered light brightness \cite{ert11}. This is
critical to better understand the implications of our HOSTS results (obtained in the mid-infrared) for exo-Earth imaging (to
be achieved in the visible). 

\begin{figure}[t]
 \centering
 \includegraphics[width=0.6\textwidth]{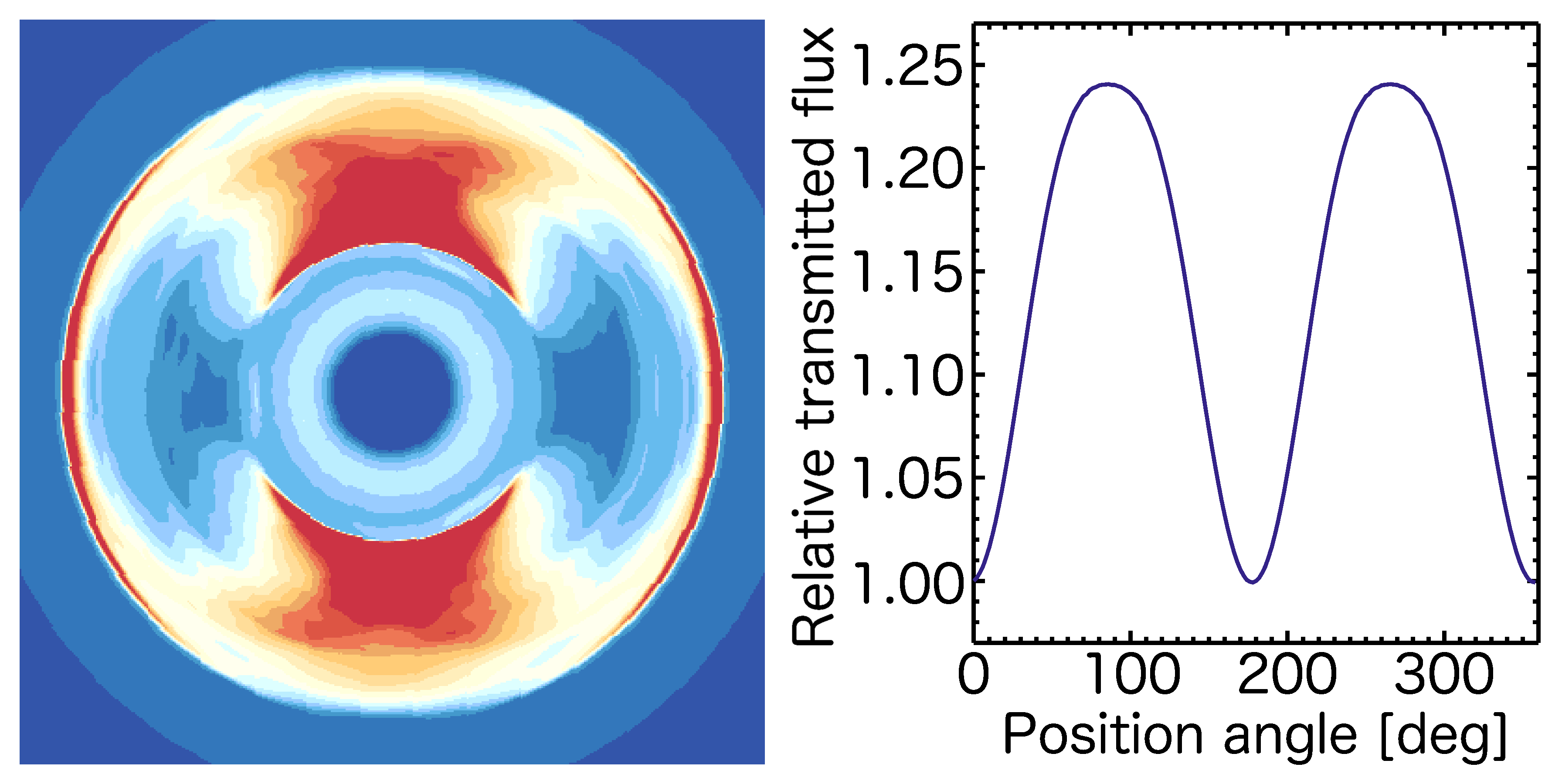}
 \caption{Simulation of observed null leak vs.\ hour angle (\textit{right}) for an observation of a disk with resonant
   structures due to planet-disk-interaction (\textit{left}, \cite{sha15}). The dust surface density is shown in the left
   image, where red corresponds to high and blue to low surface density. Simulations have been performed for a 20\,M$_\oplus$
   planet at a separation of 1\,au from a Sun-like star (distance 10\,pc).}
 \label{fig_structures}
\end{figure}

We demonstrate the feasibility of the proposed observations using simulations of the expected signal for several scenarios,
using our detection around $\beta$\,UMa as a template. Fig.~\ref{fig_inclinations} shows the null excess vs.\ hour angle of
observations for two disks at different inclinations and position angles. It can be seen that the null excess varies by
$\sim$20\% and $\sim$35\% for the two examples due to the disk inclination. Fig.~\ref{fig_structures} shows the structure
produced by a 20\,M$_\oplus$ planet at 1\,AU from a Sun-like star in dust spiraling inward from an asteroid belt at 2.25\,AU
due to Poynting-Robertson drag \cite{sha15} and the expected null variation vs.\ position angle of LBTI's transmission
pattern. Again, a variation of $\sim$20\% can be expected. LBTI's current accuracy to null excess is $\sim6\times10^{-4}$
(1\,$\sigma$) for a calibrated observing sequence. Thus, the expected signals can be detected with two observations in the
extrema of the null variation if the mean excess is $\sim$1\%. They should thus be well detectable in our systems with the
strongest excesses ($\beta$\,Leo, $\beta$\,UMa, $\eta$\,Crv, $\zeta$\,Lep, 13\,UMa). In addition, strong structures on top of
the measured excesses may be visible for our other detections including $\epsilon$\,Eri and Vega. Because the hour angles of
the extrema of the null excess are unknown a priori, observations need to be carried out at a range of hour angles. This will
naturally produce more data than observations in the extrema only and thus yield a higher sensitivity to the observed
features. A realistic observing cadence and hour angle coverage of observations optimized to detect variations of the null
excess with hour angle is shown in Fig.~\ref{fig_inclinations}.

\begin{figure}[t]
 \centering
 \includegraphics[width=1.0\textwidth]{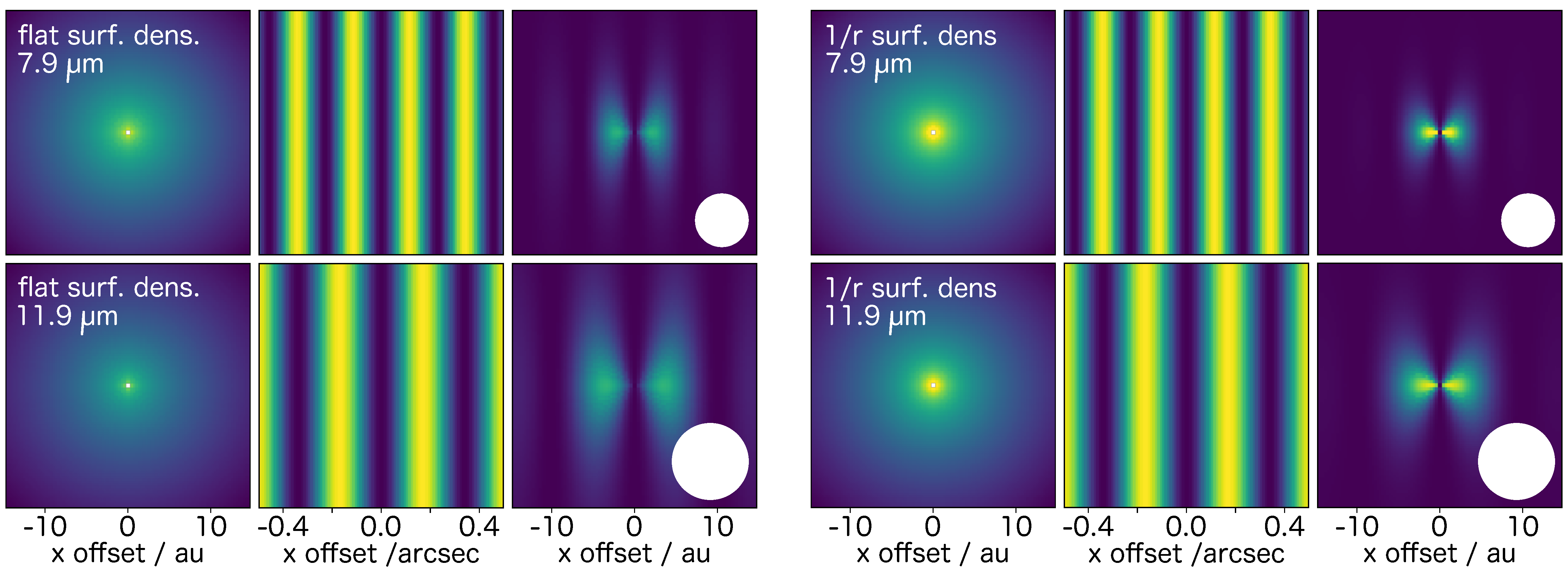}
 \caption{Transmission of disk flux at 7.9\,$\mu$m (\textit{top row}) and 11.9\,$\mu$m (\textit{bottom row}) for two disk
  models with different radial surface density (\textit{left three columns}: flat, \textit{right three}: $1/r$). The disk
  model image (in logarithmic scale), transmission pattern, and transmitted flux are shown. All disk model images are shown
  in the same color scale. Transmitted flux is shown in the same color scale for the same wavelength but the peak brightness
  is twice as high at 7.9\,$\mu$m than at 11.9\,$\mu$m. The flux ratio of the null excess at 11.9\,$\mu$m and 7.9\,$\mu$m is
  $N_{11.9}/N_{7.9} = 1.4$ for the flat surface density and $N_{11.9}/N_{7.9} = 0.9$ for the $1/r$ case. The white circles
  illustrate the FWHM of the single telescope PSF. It can be seen that the transmitted flux is resolved. Our detection of
  $\beta$\,UMa is used as a template.}
 \label{fig_wavelengths}
\end{figure}

Simulations at two wavelengths (e.g., 7.9\,$\mu$m and 11.9\,$\mu$m, filter width $\sim$10\%) of a face-on disk with two
different surface density slopes (flat and outward decreasing with $1/r$) are shown in Fig.~\ref{fig_wavelengths}. From these
data one can measure the null excess with different photometric aperture sizes to reconstruct its angular size which is at
least marginally resolved by the single aperture PSF. Moreover, one can measure the null excess ratio between two
wavelengths, which is $N_{11.9}/N_{7.9} = 1.4$ and $N_{11.9}/N_{7.9} = 0.9$ for the flat and $1/r$ surface density cases,
respectively. This difference is also detectable for null excesses around 1\% with single observations in each wavelength,
and for lower excesses if more data are obtained.

Detailed model fitting, combining the wealth of data provided by the observations described above can also break degeneracies
in single data sets \cite{ken15, ert11, ert12a, ert14a}. Such degeneracies are for example present between disk inclination
and structure as a source of hour angle variation of the measured excess or between dust grain size and radial distribution
in the wavelength dependence of the measured null excesses. However, the first can be broken when measuring the variation at
two wavelengths (the inclination effect is related to the inner working angle, the effect of a clump typically not). The
latter can be broken by measuring the radial distribution also with different photometric aperture sizes.

\subsection{Characterization of detected systems using Fizeau spectro-interferometry}
Another possibility the LBTI offers for the characterization of exozodiacal dust is spectro-interferometry in the $K$~to
$N$~band in constructive Fizeau mode. The LBTI is currently the only interferometer operating in this wavelength range and
using a fringe tracker. Thus, it should in principle be possible to obtain spectrally resolved data with an accuracy
sufficient to detect excesses at the $\sim$1\% level. Such data could for example be used to study the connection between HZ
dust and hot dust closer in by tracing the warm dust emission toward shorter wavelengths of the hot emission toward longer
wavelengths. It should also be possible to detect the 3\,$\mu$m and 10\,$\mu$m silicate features of the dust to further study
its composition. A description of LBTI's Fizeau mode and our efforts in implementing routine, accurate, and efficient
observations is presented by \cite{spa18}.

\section{Conclusions}
\label{sect_conc}

We have conducted the most sensitive survey for HZ dust around nearby stars. We put strong constraints on the typical HZ dust
levels and show that these are low enough for stars without known cold dust to allow exo-Earth imaging. However, our
detections around these stars demonstrate that such a target selection strategy does not guarantee low HZ dust levels for all
stars. This limits our constraint on the median dust level to 16~zodis, while for a sample of LBTI vetted stars the median
level would be by a factor of two better. This transforms into a 20\% increase of the yield of an exo-Earth imaging mission
of a given size. We have also outlined the science case for a future characterization of the HOSTS detected exozodis with the
LBTI and demonstrated the feasibility of the proposed observations through modeling of the expected signals. These
observations represent a critical next step in the understanding of the origin and properties of exozodiacal dust and provide
present-day insights into the architecture and dynamics of nearby planetary systems near their HZs.

\acknowledgments

The Large Binocular Telescope Interferometer is funded by the National Aeronautics and Space Administration as part of
its Exoplanet Exploration Program. The LBT is an international collaboration among institutions in the United States,
Italy, and Germany. LBT Corporation partners are: The University of Arizona on behalf of the Arizona university system;
Instituto Nazionale di Astrofisica, Italy; LBT Beteiligungsgesellschaft, Germany, representing the Max-Planck Society,
the Astrophysical Institute Potsdam, and Heidelberg University; The Ohio State University, and The Research
Corporation, on behalf of The University of Notre Dame, University of Minnesota and University of Virginia. This
research has made extensive use of the SIMBAD database \cite{wen00} and the VizieR catalogue access tool
\cite{och00}, both operated at CDS, Strasbourg, France, of Python, including the NumPy, SciPy, Matplotlib
\cite{hun07}, and Astorpy \cite{ast13} libraries, and of NASA's Astrophysics Data System Bibliographic Services. GMK
is supported by the Royal Society as a Royal Society University Research Fellow. AS is partially supported by funding
from the Center for Exoplanets and Habitable Worlds. The Center for Exoplanets and Habitable Worlds is supported by the
Pennsylvania State University, the Eberly College of Science, and the Pennsylvania Space Grant Consortium. JMS is supported
by NASA through Hubble Fellowship grant HST-HF2-51398.001-A awarded by the Space Telescope Science Institute, which is
operated by the Association of Universities for Research in Astronomy, Inc., for NASA, under contract NAS5-26555.

\bibliography{../../../bibtex} 

\begin{thebibliography}{10}

\bibitem{mil11}
{Millan-Gabet}, R., {Serabyn}, E., {Mennesson}, B., {Stark}, C.~C., {Ragland},
  S., {Hrynevych}, M., {Woillez}, J., {Stapelfeldt}, K., {Bryden}, G.,
  {Colavita}, M.~M., and {Booth}, A.~J., ``{Exozodiacal Dust Levels for Nearby
  Main-sequence Stars: A Survey with the Keck Interferometer Nuller},'' {\em
  \apj}~{\bf 734},  67 (June 2011).

\bibitem{men14}
{Mennesson}, B., {Millan-Gabet}, R., {Serabyn}, E., {Colavita}, M.~M., {Absil},
  O., {Bryden}, G., {Wyatt}, M., {Danchi}, W., {Defr{\`e}re}, D., {Dor{\'e}},
  O., {Hinz}, P., {Kuchner}, M., {Ragland}, S., {Scott}, N., {Stapelfeldt}, K.,
  {Traub}, W., and {Woillez}, J., ``{Constraining the Exozodiacal Luminosity
  Function of Main-sequence Stars: Complete Results from the Keck Nuller
  Mid-infrared Surveys},'' {\em \apj}~{\bf 797},  119 (Dec. 2014).

\bibitem{ert18}
{Ertel}, S., {Defr{\`e}re}, D., {Hinz}, P., {Mennesson}, B., {Kennedy}, G.~M.,
  {Danchi}, W.~C., {Gelino}, C., {Hill}, J.~M., {Hoffmann}, W.~F., {Rieke}, G.,
  {Shannon}, A., {Spalding}, E., {Stone}, J.~M., {Vaz}, A., {Weinberger},
  A.~J., {Willems}, P., {Absil}, O., {Arbo}, P., {Bailey}, V.~P., {Beichman},
  C., {Bryden}, G., {Downey}, E.~C., {Durney}, O., {Esposito}, S., {Gaspar},
  A., {Grenz}, P., {Haniff}, C.~A., {Leisenring}, J.~M., {Marion}, L.,
  {McMahon}, T.~J., {Millan-Gabet}, R., {Montoya}, M., {Morzinski}, K.~M.,
  {Pinna}, E., {Power}, J., {Puglisi}, A., {Roberge}, A., {Serabyn}, E.,
  {Skemer}, A.~J., {Stapelfeldt}, K., {Su}, K.~Y.~L., {Vaitheeswaran}, V., and
  {Wyatt}, M.~C., ``{The HOSTS Survey -- Exozodiacal Dust Measurements for 30
  Stars},'' {\em \aj}~{\bf 155},  194 (May 2018).

\bibitem{abs06}
{Absil}, O., {di Folco}, E., {M{\'e}rand}, A., {Augereau}, J.-C., {Coud{\'e} du
  Foresto}, V., {Aufdenberg}, J.~P., {Kervella}, P., {Ridgway}, S.~T.,
  {Berger}, D.~H., {ten Brummelaar}, T.~A., {Sturmann}, J., {Sturmann}, L.,
  {Turner}, N.~H., and {McAlister}, H.~A., ``{Circumstellar material in the
  <ASTROBJ>Vega</ASTROBJ> inner system revealed by CHARA/FLUOR},'' {\em
  \aap}~{\bf 452},  237--244 (June 2006).

\bibitem{abs13}
{Absil}, O., {Defr{\`e}re}, D., {Coud{\'e} du Foresto}, V., {Di Folco}, E.,
  {M{\'e}rand}, A., {Augereau}, J.-C., {Ertel}, S., {Hanot}, C., {Kervella},
  P., {Mollier}, B., {Scott}, N., {Che}, X., {Monnier}, J.~D., {Thureau}, N.,
  {Tuthill}, P.~G., {ten Brummelaar}, T.~A., {McAlister}, H.~A., {Sturmann},
  J., {Sturmann}, L., and {Turner}, N., ``{A near-infrared interferometric
  survey of debris-disc stars. III. First statistics based on 42 stars observed
  with CHARA/FLUOR},'' {\em \aap}~{\bf 555},  A104 (July 2013).

\bibitem{ert14b}
{Ertel}, S., {Absil}, O., {Defr{\`e}re}, D., {Le Bouquin}, J.-B., {Augereau},
  J.-C., {Marion}, L., {Blind}, N., {Bonsor}, A., {Bryden}, G., {Lebreton}, J.,
  and {Milli}, J., ``{A near-infrared interferometric survey of debris-disk
  stars. IV. An unbiased sample of 92 southern stars observed in H band with
  VLTI/PIONIER},'' {\em \aap}~{\bf 570},  A128 (Oct. 2014).

\bibitem{ert16}
{Ertel}, S., {Defr{\`e}re}, D., {Absil}, O., {Le Bouquin}, J.-B., {Augereau},
  J.-C., {Berger}, J.-P., {Blind}, N., {Bonsor}, A., {Lagrange}, A.-M.,
  {Lebreton}, J., {Marion}, L., {Milli}, J., and {Olofsson}, J., ``{A
  near-infrared interferometric survey of debris-disc stars. V. PIONIER search
  for variability},'' {\em \aap}~{\bf 595},  A44 (Oct. 2016).

\bibitem{kra17}
{Kral}, Q., {Krivov}, A.~V., {Defr{\`e}re}, D., {van Lieshout}, R., {Bonsor},
  A., {Augereau}, J.-C., {Th{\'e}bault}, P., {Ertel}, S., {Lebreton}, J., and
  {Absil}, O., ``{Exozodiacal clouds: hot and warm dust around main sequence
  stars},'' {\em The Astronomical Review}~{\bf 13},  69--111 (Apr. 2017).

\bibitem{far17}
{Faramaz}, V., {Ertel}, S., {Booth}, M., {Cuadra}, J., and {Simmonds}, C.,
  ``{Inner mean-motion resonances with eccentric planets: a possible origin for
  exozodiacal dust clouds},'' {\em \mnras}~{\bf 465},  2352--2365 (Feb. 2017).

\bibitem{bac93}
{Backman}, D.~E. and {Paresce}, F., ``{Main-sequence stars with circumstellar
  solid material - The VEGA phenomenon},'' in [{\em Protostars and Planets
  III}{\nolinebreak\hspace{0.1em}]},  {E.~H.~Levy \& J.~I.~Lunine}, ed.,
  1253--1304 (1993).

\bibitem{ken15b}
{Kennedy}, G.~M. and {Piette}, A., ``{Warm exo-Zodi from cool exo-Kuiper belts:
  the significance of P-R drag and the inference of intervening planets},''
  {\em \mnras}~{\bf 449},  2304--2311 (May 2015).

\bibitem{rei11}
{Reidemeister}, M., {Krivov}, A.~V., {Stark}, C.~C., {Augereau}, J.,
  {L{\"o}hne}, T., and {M{\"u}ller}, S., ``{The cold origin of the warm dust
  around {$\epsilon$} Eridani},'' {\em \aap}~{\bf 527},  A57+ (Mar. 2011).

\bibitem{ert12b}
{Ertel}, S., {Wolf}, S., and {Rodmann}, J., ``{Observing planet-disk
  interaction in debris disks},'' {\em \aap}~{\bf 544},  A61 (Aug. 2012).

\bibitem{sha15}
{Shannon}, A., {Mustill}, A.~J., and {Wyatt}, M., ``{Capture and evolution of
  dust in planetary mean-motion resonances: a fast, semi-analytic method for
  generating resonantly trapped disc images},'' {\em \mnras}~{\bf 448},
  684--702 (Mar. 2015).

\bibitem{sta15}
{Stark}, C.~C., {Roberge}, A., {Mandell}, A., {Clampin}, M., {Domagal-Goldman},
  S.~D., {McElwain}, M.~W., and {Stapelfeldt}, K.~R., ``{Lower Limits on
  Aperture Size for an ExoEarth Detecting Coronagraphic Mission},'' {\em
  \apj}~{\bf 808},  149 (Aug. 2015).

\bibitem{sta16}
{Stark}, C.~C., {Shaklan}, S., {Lisman}, D., {Cady}, E., {Savransky}, D.,
  {Roberge}, A., and {Mandell}, A.~M., ``{Maximized exoEarth candidate yields
  for starshades},'' {\em Journal of Astronomical Telescopes, Instruments, and
  Systems}~{\bf 2},  041204 (Oct. 2016).

\bibitem{def10}
{Defr{\`e}re}, D., {Absil}, O., {den Hartog}, R., {Hanot}, C., and {Stark}, C.,
  ``{Nulling interferometry: impact of exozodiacal clouds on the performance of
  future life-finding space missions},'' {\em \aap}~{\bf 509},  A9 (Jan. 2010).

\bibitem{hin16}
{Hinz}, P.~M., {Defr{\`e}re}, D., {Skemer}, A., {Bailey}, V., {Stone}, J.,
  {Spalding}, E., {Vaz}, A., {Pinna}, E., {Puglisi}, A., {Esposito}, S.,
  {Montoya}, M., {Downey}, E., {Leisenring}, J., {Durney}, O., {Hoffmann}, W.,
  {Hill}, J., {Millan-Gabet}, R., {Mennesson}, B., {Danchi}, W., {Morzinski},
  K., {Grenz}, P., {Skrutskie}, M., and {Ertel}, S., ``{Overview of LBTI: a
  multipurpose facility for high spatial resolution observations},'' in [{\em
  Optical and Infrared Interferometry and Imaging
  V}{\nolinebreak\hspace{0.1em}]},  {\em \procspie} {\bf 9907},  990704 (Aug.
  2016).

\bibitem{wei15}
{Weinberger}, A.~J., {Bryden}, G., {Kennedy}, G.~M., {Roberge}, A.,
  {Defr{\`e}re}, D., {Hinz}, P.~M., {Millan-Gabet}, R., {Rieke}, G., {Bailey},
  V.~P., {Danchi}, W.~C., {Haniff}, C., {Mennesson}, B., {Serabyn}, E.,
  {Skemer}, A.~J., {Stapelfeldt}, K.~R., and {Wyatt}, M.~C., ``{Target
  Selection for the LBTI Exozodi Key Science Program},'' {\em \apjs}~{\bf 216},
   24 (Feb. 2015).

\bibitem{bor02}
{Bord{\'e}}, P., {Coud{\'e} du Foresto}, V., {Chagnon}, G., and {Perrin}, G.,
  ``{A catalogue of calibrator stars for long baseline stellar
  interferometry},'' {\em \aap}~{\bf 393},  183--193 (Oct. 2002).

\bibitem{mer05}
{M{\'e}rand}, A., {Bord{\'e}}, P., and {Coud{\'e} du Foresto}, V., ``{A catalog
  of bright calibrator stars for 200-m baseline near-infrared stellar
  interferometry},'' {\em \aap}~{\bf 433},  1155--1162 (Apr. 2005).

\bibitem{che16}
{Chelli}, A., {Duvert}, G., {Bourg{\`e}s}, L., {Mella}, G., {Lafrasse}, S.,
  {Bonneau}, D., and {Chesneau}, O., ``{Pseudomagnitudes and differential
  surface brightness: Application to the apparent diameter of stars},'' {\em
  \aap}~{\bf 589},  A112 (May 2016).

\bibitem{gas13}
{G{\'a}sp{\'a}r}, A., {Rieke}, G.~H., and {Balog}, Z., ``{The Collisional
  Evolution of Debris Disks},'' {\em \apj}~{\bf 768},  25 (May 2013).

\bibitem{thu14}
{Thureau}, N.~D., {Greaves}, J.~S., {Matthews}, B.~C., {Kennedy}, G.,
  {Phillips}, N., {Booth}, M., {Duch{\^e}ne}, G., {Horner}, J., {Rodriguez},
  D.~R., {Sibthorpe}, B., and {Wyatt}, M.~C., ``{An unbiased study of debris
  discs around A-type stars with Herschel},'' {\em \mnras}~{\bf 445},
  2558--2573 (Dec. 2014).

\bibitem{bei06}
{Beichman}, C.~A., {Bryden}, G., {Stapelfeldt}, K.~R., {Gautier}, T.~N.,
  {Grogan}, K., {Shao}, M., {Velusamy}, T., {Lawler}, S.~M., {Blaylock}, M.,
  {Rieke}, G.~H., {Lunine}, J.~I., {Fischer}, D.~A., {Marcy}, G.~W., {Greaves},
  J.~S., {Wyatt}, M.~C., {Holland}, W.~S., and {Dent}, W.~R.~F., ``{New Debris
  Disks around Nearby Main-Sequence Stars: Impact on the Direct Detection of
  Planets},'' {\em \apj}~{\bf 652},  1674--1693 (Dec. 2006).

\bibitem{su06}
{Su}, K.~Y.~L., {Rieke}, G.~H., {Stansberry}, J.~A., {Bryden}, G.,
  {Stapelfeldt}, K.~R., {Trilling}, D.~E., {Muzerolle}, J., {Beichman}, C.~A.,
  {Moro-Martin}, A., {Hines}, D.~C., and {Werner}, M.~W., ``{Debris Disk
  Evolution around A Stars},'' {\em \apj}~{\bf 653},  675--689 (Dec. 2006).

\bibitem{rie05}
{Rieke}, G.~H., {Su}, K.~Y.~L., {Stansberry}, J.~A., {Trilling}, D., {Bryden},
  G., {Muzerolle}, J., {White}, B., {Gorlova}, N., {Young}, E.~T., {Beichman},
  C.~A., {Stapelfeldt}, K.~R., and {Hines}, D.~C., ``{Decay of Planetary Debris
  Disks},'' {\em \apj}~{\bf 620},  1010--1026 (Feb. 2005).

\bibitem{eir13}
{Eiroa}, C., {Marshall}, J.~P., {Mora}, A., {Montesinos}, B., {Absil}, O.,
  {Augereau}, J.~C., {Bayo}, A., {Bryden}, G., {Danchi}, W., {del Burgo}, C.,
  {Ertel}, S., {Fridlund}, M., {Heras}, A.~M., {Krivov}, A.~V., {Launhardt},
  R., {Liseau}, R., {L{\"o}hne}, T., {Maldonado}, J., {Pilbratt}, G.~L.,
  {Roberge}, A., {Rodmann}, J., {Sanz-Forcada}, J., {Solano}, E.,
  {Stapelfeldt}, K., {Th{\'e}bault}, P., {Wolf}, S., {Ardila}, D.,
  {Ar{\'e}valo}, M., {Beichmann}, C., {Faramaz}, V.,
  {Gonz{\'a}lez-Garc{\'{\i}}a}, B.~M., {Guti{\'e}rrez}, R., {Lebreton}, J.,
  {Mart{\'{\i}}nez-Arn{\'a}iz}, R., {Meeus}, G., {Montes}, D., {Olofsson}, G.,
  {Su}, K.~Y.~L., {White}, G.~J., {Barrado}, D., {Fukagawa}, M., {Gr{\"u}n},
  E., {Kamp}, I., {Lorente}, R., {Morbidelli}, A., {M{\"u}ller}, S.,
  {Mutschke}, H., {Nakagawa}, T., {Ribas}, I., and {Walker}, H., ``{DUst around
  NEarby Stars. The survey observational results},'' {\em \aap}~{\bf 555},  A11
  (July 2013).

\bibitem{mon16}
{Montesinos}, B., {Eiroa}, C., {Krivov}, A.~V., {Marshall}, J.~P., {Pilbratt},
  G.~L., {Liseau}, R., {Mora}, A., {Maldonado}, J., {Wolf}, S., {Ertel}, S.,
  {Bayo}, A., {Augereau}, J.-C., {Heras}, A.~M., {Fridlund}, M., {Danchi},
  W.~C., {Solano}, E., {Kirchschlager}, F., {del Burgo}, C., and {Montes}, D.,
  ``{Incidence of debris discs around FGK stars in the solar neighbourhood},''
  {\em \aap}~{\bf 593},  A51 (Sept. 2016).

\bibitem{tri08}
{Trilling}, D.~E., {Bryden}, G., {Beichman}, C.~A., {Rieke}, G.~H., {Su},
  K.~Y.~L., {Stansberry}, J.~A., {Blaylock}, M., {Stapelfeldt}, K.~R.,
  {Beeman}, J.~W., and {Haller}, E.~E., ``{Debris Disks around Sun-like
  Stars},'' {\em \apj}~{\bf 674},  1086--1105 (Feb. 2008).

\bibitem{gre04}
{Greaves}, J.~S., {Wyatt}, M.~C., {Holland}, W.~S., and {Dent}, W.~R.~F.,
  ``{The debris disc around {\ensuremath{\tau}} Ceti: a massive analogue to the
  Kuiper Belt},'' {\em \mnras}~{\bf 351},  L54--L58 (July 2004).

\bibitem{aum85}
{Aumann}, H.~H., ``{IRAS observations of matter around nearby stars},'' {\em
  \pasp}~{\bf 97},  885--891 (Oct. 1985).

\bibitem{law09}
{Lawler}, S.~M., {Beichman}, C.~A., {Bryden}, G., {Ciardi}, D.~R., {Tanner},
  A.~M., {Su}, K.~Y.~L., {Stapelfeldt}, K.~R., {Lisse}, C.~M., and {Harker},
  D.~E., ``{Explorations Beyond the Snow Line: Spitzer/IRS Spectra of Debris
  Disks Around Solar-type Stars},'' {\em \apj}~{\bf 705},  89--111 (Nov. 2009).

\bibitem{koe10}
{Koerner}, D.~W., {Kim}, S., {Trilling}, D.~E., {Larson}, H., {Cotera}, A.,
  {Stapelfeldt}, K.~R., {Wahhaj}, Z., {Fajardo-Acosta}, S., {Padgett}, D., and
  {Backman}, D., ``{New Debris Disk Candidates Around 49 Nearby Stars},'' {\em
  \apjl}~{\bf 710},  L26--L29 (Feb. 2010).

\bibitem{aum88}
{Aumann}, H.~H., ``{Spectral class distribution of circumstellar material in
  main-sequence stars},'' {\em \aj}~{\bf 96},  1415--1419 (Oct. 1988).

\bibitem{kha07}
{Kharchenko}, N.~V., {Scholz}, R., {Piskunov}, A.~E., {Roeser}, S., and
  {Schilbach}, E., ``{2nd Cat. of Radial Velocities with Astrometric Data
  (Kharchenko+, 2007)},'' {\em VizieR Online Data Catalog}~{\bf 3254},  0--+
  (June 2007).

\bibitem{gez93}
{Gezari}, D.~Y., {Schmitz}, M., {Pitts}, P.~S., and {Mead}, J.~M.,  [{\em
  {Catalog of infrared observations, third
  edition}}{\nolinebreak\hspace{0.1em}]} (June 1993).

\bibitem{vanle07}
{van Leeuwen}, F., ``{Validation of the new Hipparcos reduction},'' {\em
  \aap}~{\bf 474},  653--664 (Nov. 2007).

\bibitem{hof14}
{Hoffmann}, W.~F., {Hinz}, P.~M., {Defr{\`e}re}, D., {Leisenring}, J.~M.,
  {Skemer}, A.~J., {Arbo}, P.~A., {Montoya}, M., and {Mennesson}, B.,
  ``{Operation and performance of the mid-infrared camera, NOMIC, on the Large
  Binocular Telescope},'' in [{\em Ground-based and Airborne Instrumentation
  for Astronomy V}{\nolinebreak\hspace{0.1em}]},  {\em \procspie} {\bf 9147},
  91471O (July 2014).

\bibitem{def15b}
{Defr{\`e}re}, D., {Hinz}, P., {Downey}, E., {Ashby}, D., {Bailey}, V.,
  {Brusa}, G., {Christou}, J., {Danchi}, W.~C., {Grenz}, P., {Hill}, J.~M.,
  {Hoffmann}, W.~F., {Leisenring}, J., {Lozi}, J., {McMahon}, T., {Mennesson},
  B., {Millan-Gabet}, R., {Montoya}, M., {Powell}, K., {Skemer}, A.,
  {Vaitheeswaran}, V., {Vaz}, A., and {Veillet}, C., ``{Co-phasing the Large
  Binocular Telescope: status and performance of LBTI/PHASECam},'' {\em ArXiv
  e-prints}  (Jan. 2015).

\bibitem{men11}
{Mennesson}, B., {Serabyn}, E., {Hanot}, C., {Martin}, S.~R., {Liewer}, K., and
  {Mawet}, D., ``{New Constraints on Companions and Dust within a Few AU of
  Vega},'' {\em \apj}~{\bf 736},  14 (July 2011).

\bibitem{han11}
{Hanot}, C., {Mennesson}, B., {Martin}, S., {Liewer}, K., {Loya}, F., {Mawet},
  D., {Riaud}, P., {Absil}, O., and {Serabyn}, E., ``{Improving Interferometric
  Null Depth Measurements using Statistical Distributions: Theory and First
  Results with the Palomar Fiber Nuller},'' {\em \apj}~{\bf 729},  110 (Mar.
  2011).

\bibitem{def16}
{Defr{\`e}re}, D., {Hinz}, P.~M., {Mennesson}, B., {Hoffmann}, W.~F.,
  {Millan-Gabet}, R., {Skemer}, A.~J., {Bailey}, V., {Danchi}, W.~C., {Downey},
  E.~C., {Durney}, O., {Grenz}, P., {Hill}, J.~M., {McMahon}, T.~J., {Montoya},
  M., {Spalding}, E., {Vaz}, A., {Absil}, O., {Arbo}, P., {Bailey}, H.,
  {Brusa}, G., {Bryden}, G., {Esposito}, S., {Gaspar}, A., {Haniff}, C.~A.,
  {Kennedy}, G.~M., {Leisenring}, J.~M., {Marion}, L., {Nowak}, M., {Pinna},
  E., {Powell}, K., {Puglisi}, A., {Rieke}, G., {Roberge}, A., {Serabyn}, E.,
  {Sosa}, R., {Stapeldfeldt}, K., {Su}, K., {Weinberger}, A.~J., and {Wyatt},
  M.~C., ``{Nulling Data Reduction and On-sky Performance of the Large
  Binocular Telescope Interferometer},'' {\em \apj}~{\bf 824},  66 (June 2016).

\bibitem{men16a}
{Mennesson}, B., {Defr{\`e}re}, D., {Nowak}, M., {Hinz}, P., {Millan-Gabet},
  R., {Absil}, O., {Bailey}, V., {Bryden}, G., {Danchi}, W., {Kennedy}, G.~M.,
  {Marion}, L., {Roberge}, A., {Serabyn}, E., {Skemer}, A.~J., {Stapelfeldt},
  K., {Weinberger}, A.~J., and {Wyatt}, M., ``{Making high-accuracy null depth
  measurements for the LBTI exozodi survey},'' in [{\em Optical and Infrared
  Interferometry and Imaging V}{\nolinebreak\hspace{0.1em}]},  {\em \procspie}
  {\bf 9907},  99070X (Aug. 2016).

\bibitem{kel98}
{Kelsall}, T., {Weiland}, J.~L., {Franz}, B.~A., {Reach}, W.~T., {Arendt},
  R.~G., {Dwek}, E., {Freudenreich}, H.~T., {Hauser}, M.~G., {Moseley}, S.~H.,
  {Odegard}, N.~P., {Silverberg}, R.~F., and {Wright}, E.~L., ``{The COBE
  Diffuse Infrared Background Experiment Search for the Cosmic Infrared
  Background. II. Model of the Interplanetary Dust Cloud},'' {\em \apj}~{\bf
  508},  44--73 (Nov. 1998).

\bibitem{ken15}
{Kennedy}, G.~M., {Wyatt}, M.~C., {Bailey}, V., {Bryden}, G., {Danchi}, W.~C.,
  {Defr{\`e}re}, D., {Haniff}, C., {Hinz}, P.~M., {Lebreton}, J., {Mennesson},
  B., {Millan-Gabet}, R., {Morales}, F., {Pani{\'c}}, O., {Rieke}, G.~H.,
  {Roberge}, A., {Serabyn}, E., {Shannon}, A., {Skemer}, A.~J., {Stapelfeldt},
  K.~R., {Su}, K.~Y.~L., and {Weinberger}, A.~J., ``{Exo-zodi Modeling for the
  Large Binocular Telescope Interferometer},'' {\em \apjs}~{\bf 216},  23 (Feb.
  2015).

\bibitem{law14}
{Lawler}, S.~M., {Di Francesco}, J., {Kennedy}, G.~M., {Sibthorpe}, B.,
  {Booth}, M., {Vandenbussche}, B., {Matthews}, B.~C., {Holland}, W.~S.,
  {Greaves}, J., {Wilner}, D.~J., {Tuomi}, M., {Blommaert}, J.~A.~D.~L., {de
  Vries}, B.~L., {Dominik}, C., {Fridlund}, M., {Gear}, W., {Heras}, A.~M.,
  {Ivison}, R., and {Olofsson}, G., ``{The debris disc of solar analogue
  {\ensuremath{\tau}} Ceti: Herschel observations and dynamical simulations of
  the proposed multiplanet system},'' {\em \mnras}~{\bf 444},  2665--2675 (Nov.
  2014).

\bibitem{fen17}
{Feng}, F., {Tuomi}, M., {Jones}, H.~R.~A., {Barnes}, J., {Anglada-Escud{\'e}},
  G., {Vogt}, S.~S., and {Butler}, R.~P., ``{Color Difference Makes a
  Difference: Four Planet Candidates around {$\tau$} Ceti},'' {\em \aj}~{\bf
  154},  135 (Oct. 2017).

\bibitem{def18}
{Defr{\`e}re}, D., {Absil}, O., {Berger}, J.-P., {Boulet}, T., {Danchi}, W.~C.,
  {Ertel}, S., {Gallenne}, A., {H{\'e}nault}, F., {Hinz}, P., {Huby}, E.,
  {Ireland}, M., {Kraus}, S., {Labadie}, L., {Le Bouquin}, J.-B., {Martin}, G.,
  {Matter}, A., {M{\'e}rand}, A., {Mennesson}, B., {Minardi}, S., {Monnier},
  J., {Norris}, B., {Orban de Xivry}, G., {Pedretti}, E., {Pott}, J.-U.,
  {Reggiani}, M., {Serabyn}, E., {Surdej}, J., {Tristram}, K.~R.~W., and
  {Woillez}, J., ``{The path towards high-contrast imaging with the VLTI: the
  Hi-5 project},'' {\em ArXiv e-prints}  (Jan. 2018).

\bibitem{def18b}
{Defr{\`e}re}, D., {Absil}, O., {Berger}, J.-P., {Boulet}, T., {Danchi}, W.~C.,
  {Ertel}, S., {Gallenne}, A., {H{\'e}nault}, F., {Hinz}, P., {Huby}, E.,
  {Ireland}, M., {Kraus}, S., {Labadie}, L., {Le Bouquin}, J.-B., {Martin}, G.,
  {Matter}, A., {M{\'e}rand}, A., {Mennesson}, B., {Minardi}, S., {Monnier},
  J., {Norris}, B., {Orban de Xivry}, G., {Pedretti}, E., {Pott}, J.-U.,
  {Reggiani}, M., {Serabyn}, E., {Surdej}, J., {Tristram}, K.~R.~W., and
  {Woillez}, J., ``{Hi-5: a potential high-contrast thermal near-infrared
  imager for the VLTI},'' To appear in the same issue as the present paper.
  (2018).

\bibitem{nes10}
{Nesvorn{\'y}}, D., {Jenniskens}, P., {Levison}, H.~F., {Bottke}, W.~F.,
  {Vokrouhlick{\'y}}, D., and {Gounelle}, M., ``{Cometary Origin of the
  Zodiacal Cloud and Carbonaceous Micrometeorites. Implications for Hot Debris
  Disks},'' {\em \apj}~{\bf 713},  816--836 (Apr. 2010).

\bibitem{mar16}
{Marshall}, J.~P., {Cotton}, D.~V., {Bott}, K., {Ertel}, S., {Kennedy}, G.~M.,
  {Wyatt}, M.~C., {del Burgo}, C., {Absil}, O., {Bailey}, J., and
  {Kedziora-Chudczer}, L., ``{Polarization Measurements of Hot Dust Stars and
  the Local Interstellar Medium},'' {\em \apj}~{\bf 825},  124 (July 2016).

\bibitem{ert12a}
{Ertel}, S., {Wolf}, S., {Marshall}, J.~P., {Eiroa}, C., {Augereau}, J.-C.,
  {Krivov}, A.~V., {L{\"o}hne}, T., {Absil}, O., {Ardila}, D., {Ar{\'e}valo},
  M., {Bayo}, A., {Bryden}, G., {del Burgo}, C., {Greaves}, J., {Kennedy}, G.,
  {Lebreton}, J., {Liseau}, R., {Maldonado}, J., {Montesinos}, B., {Mora}, A.,
  {Pilbratt}, G.~L., {Sanz-Forcada}, J., {Stapelfeldt}, K., and {White}, G.~J.,
  ``{A peculiar class of debris disks from Herschel/DUNES. A steep fall off in
  the far infrared},'' {\em \aap}~{\bf 541},  A148 (May 2012).

\bibitem{ert11}
{Ertel}, S., {Wolf}, S., {Metchev}, S., {Schneider}, G., {Carpenter}, J.~M.,
  {Meyer}, M.~R., {Hillenbrand}, L.~A., and {Silverstone}, M.~D.,
  ``{Multi-wavelength modeling of the spatially resolved debris disk of HD
  107146},'' {\em \aap}~{\bf 533},  A132+ (Sept. 2011).

\bibitem{ert14a}
{Ertel}, S., {Marshall}, J.~P., {Augereau}, J.-C., {Krivov}, A.~V.,
  {L{\"o}hne}, T., {Eiroa}, C., {Mora}, A., {del Burgo}, C., {Montesinos}, B.,
  {Bryden}, G., {Danchi}, W., {Kirchschlager}, F., {Liseau}, R., {Maldonado},
  J., {Pilbratt}, G.~L., {Sch{\"u}ppler}, C., {Th{\'e}bault}, P., {White},
  G.~J., and {Wolf}, S., ``{Potential multi-component structure of the debris
  disk around HIP 17439 revealed by Herschel/DUNES},'' {\em \aap}~{\bf 561},
  A114 (Jan. 2014).

\bibitem{spa18}
{Spalding}, E., {Hinz}, P., {Ertel}, S., {Maier}, E., and {Stone}, J.,
  ``{Towards controlled Fizeau observations with the Large Binocular
  Telescope},'' To appear in the same issue as the present paper. (2018).

\bibitem{wen00}
{Wenger}, M., {Ochsenbein}, F., {Egret}, D., {Dubois}, P., {Bonnarel}, F.,
  {Borde}, S., {Genova}, F., {Jasniewicz}, G., {Lalo{\"e}}, S., {Lesteven}, S.,
  and {Monier}, R., ``{The SIMBAD astronomical database. The CDS reference
  database for astronomical objects},'' {\em \aaps}~{\bf 143},  9--22 (Apr.
  2000).

\bibitem{och00}
{Ochsenbein}, F., {Bauer}, P., and {Marcout}, J., ``{The VizieR database of
  astronomical catalogues},'' {\em \aaps}~{\bf 143},  23--32 (Apr. 2000).

\bibitem{hun07}
Hunter, J.~D., ``Matplotlib: A 2d graphics environment,'' {\em Computing In
  Science \& Engineering}~{\bf 9}(3),  90--95 (2007).

\bibitem{ast13}
{Astropy Collaboration}, {Robitaille}, T.~P., {Tollerud}, E.~J., {Greenfield},
  P., {Droettboom}, M., {Bray}, E., {Aldcroft}, T., {Davis}, M., {Ginsburg},
  A., {Price-Whelan}, A.~M., {Kerzendorf}, W.~E., {Conley}, A., {Crighton}, N.,
  {Barbary}, K., {Muna}, D., {Ferguson}, H., {Grollier}, F., {Parikh}, M.~M.,
  {Nair}, P.~H., {Unther}, H.~M., {Deil}, C., {Woillez}, J., {Conseil}, S.,
  {Kramer}, R., {Turner}, J.~E.~H., {Singer}, L., {Fox}, R., {Weaver}, B.~A.,
  {Zabalza}, V., {Edwards}, Z.~I., {Azalee Bostroem}, K., {Burke}, D.~J.,
  {Casey}, A.~R., {Crawford}, S.~M., {Dencheva}, N., {Ely}, J., {Jenness}, T.,
  {Labrie}, K., {Lim}, P.~L., {Pierfederici}, F., {Pontzen}, A., {Ptak}, A.,
  {Refsdal}, B., {Servillat}, M., and {Streicher}, O., ``{Astropy: A community
  Python package for astronomy},'' {\em \aap}~{\bf 558},  A33 (Oct. 2013).

\end{thebibliography}
\bibliographystyle{spiebib} 

\end{document}